\documentclass[aps,twocolumn,showpacs,preprintnumbers,nofootinbib,prd,superscriptaddress,groupedaddress,]{revtex4}
\pdfoutput=1
\usepackage{amssymb,graphicx}
\usepackage{amsmath}
\usepackage{multirow}
\usepackage{epsfig}
\usepackage[usenames]{color} 
\usepackage[export]{adjustbox}

\newcommand{\beqn}{\begin{eqnarray}}
\newcommand{\eeqn}{\end{eqnarray}}
\newcommand{\beq}{\begin{equation}}
\newcommand{\eeq}{\end{equation}}

\def\mphi{m_{\phi}}

\def\pt{\tilde{p}}
\def\rt{\tilde{\rho}}

\begin{document}

\title{Regularization of instabilities in gravity theories}
\author{Fethi M.\ Ramazano\u{g}lu}
\affiliation{Department of Physics, Ko\c{c} University, \\
Rumelifeneri Yolu, 34450 Sariyer, Istanbul, Turkey}
\date{\today}

\begin{abstract}
We investigate instabilities and their regularization in theories of gravitation. Instabilities can be beneficial since their growth often leads to prominent observable signatures which makes them especially relevant to
relatively low signal-to-noise ratio measurements such as gravitational wave detections.
An indefinitely growing instability usually renders a theory unphysical, hence a desirable instability
should also come with underlying physical machinery that stops
the growth at finite values, i.e. regularization mechanisms. The prototypical gravity theory that presents such
an instability is the spontaneous scalarization
phenomena of scalar-tensor theories, which feature a tachyonic instability. 
We identify the regularization mechanisms in this theory, and show that they can be utilized to 
regularize other instabilities as well. Namely, we present theories where spontaneous
growth is triggered by a ghost rather than a tachyon, and numerically calculate
stationary solutions of scalarized neutron stars in these theories. We speculate on the possibility of
regularizing known divergent instabilities in certain gravity theories using our findings, and discuss 
alternative theories of gravitation where regularized instabilities may be present.
Even though we study many specific examples, our main point is the recognition of
regularized instabilities as a common theme and unifying mechanism in a vast array of
gravity theories.
\end{abstract}
\maketitle

\section{Introduction}
Gravitational physics has been on the forefront of science thanks to the increasing
number of gravitational wave detections in recent years~\cite{PhysRevLett.116.061102}.
 Aside from being a novel
observational channel for astrophysics, gravitational waves also enable us to test Einstein's
general relativity (GR) for strong and dynamical fields which has not been possible before. 
This led to a resurgence of interest in alternative theories of 
gravitation~\cite{PhysRevLett.116.221101,0264-9381-32-24-243001}. Instabilities,
degrees of freedom that grow to large values, possibly to infinity, from small perturbations
play an important role in alternative theories of gravitation.
One prominent example is a tachyonic instability of the scalar field in certain scalar-tensor
theories which causes the zero scalar field solution to be unstable in the presence of neutron
stars (NSs). This instability is eventually regularized, i.e. it stops growing at finite field values,
and a stable large scalar cloud  forms around the star. This phenomena
is named \emph{spontaneous scalarization}, and is particularly interesting to gravitational
wave observations. This is due to the large scalar fields causing order-of-unity deviations from GR
that are relatively easy to detect with the limited precision of the
detectors~\cite{Will:2005va,0264-9381-32-24-243001}.

Despite their eventual regularization in spontaneous scalarization, instabilities can be
a nuisance if they are not regulated.
Generic interacting spin-$2$ field theories
are well known to possess an instability due to the existence of a ghost field which renders them
unphysical. Avoiding the ghost has only recently been possible for carefully constructed
theories~\cite{deRham:2014zqa,PhysRevLett.106.231101,Hassan2012},
and the ghost can reappear if these constructions are modified, for example when
a nonminimal matter coupling is introduced~\cite{deRham:2014fha}. 

Our main aim in this work is to identify some of the cases where
instabilities can be regularized in
theories of gravitation. First, this enables us to better understand theories with 
known regularized instabilities, and discover previously unknown cases in the vast landscape of
alternative theories of gravitation. 
Discovery of spontaneous scalarization occured more than three
decades after initial scalar-tensor theories appeared~\cite{PhysRev.124.925},
and a better understanding of the underlying physics can make further
discoveries faster and more systematic.
Second, the understanding of regularization might provide tools to regularize unwanted instabilities that
plague many theories of gravitation, such as the interacting spin-$2$ case.

We start with an analysis of spontaneous scalarization, and argue that
the type of instability, a tachyon in this case, is irrelevant to the regularization mechanism.
Nonlinear terms that become dominant as the scalar field grows is the dominant reason for
regularization, while the fact that the tachyon
can only live in a finite volume inside the star might also play a part. As an explicit
demonstration of this idea, we construct a theory where a different instability, a ghost rather than a tachyon,
leads to spontaneous scalarization. We compute NS solutions in this 
\emph{ghost-based spontaneous scalarization} theory
showing that a finite stationary scalarized star solution exists. 
We also extend these results to spontaneous vectorization~\cite{Ramazanoglu:2017xbl}.
In summary, these strongly support the idea that regularization
mechanisms of spontaneous scalarization are effective for other instabilities as well.
However, \emph{proof} of the stability of these solutions requires time evolution, including the
case of scalar ghost instabilities which we examine in most detail. 
We leave time evolution studies to future work.

In the final parts of the paper, we qualitatively discuss the possibility of using the aforementioned
regularization mechanisms to build a 
well-behaved theory of spontaneously growing spin-$2$ fields, which is the natural case after 
spontaneous scalarization and vectorization which are already known. 
This is a theory of interacting spin-$2$ fields with nonminimal matter couplings as well,
which we stated to be ill-behaved for generic cases due to the reappearance of the ghost. However, 
since the regularization mechanisms of spontaneous growth seems to act irrespective
of the nature of the instability,  we speculate that the reappeared ghost may be rendered
harmless by these same mechanisms, leading to a well-behaved theory. 

Finally, we shortly discuss further forms of instabilities that might be regularized
by similar mechanisms, especially in more general scalar-tensor theories as in Horndeski and
beyond-Horndeski theories~\cite{Horndeski1974,PhysRevLett.114.211101}. All these cases show how 
diverse theories of gravitation are connected in an unexpected way through 
the role instabilities play in them.

\section{Overview of Spontaneous Scalarization}\label{sec_st}
We start with a short explanation of spontaneous scalarization following~\cite{Ramazanoglu:2016kul},
with an emphasis on the role of the instability. The action for spontaneous scalarization
is~\cite{PhysRevLett.70.2220,Chen:2015zmx,Ramazanoglu:2016kul} 
\begin{align}\label{st_action}
 \frac{1}{16\pi} &\int dV R -  \frac{1}{16\pi} \int dV \left[ 2g^{\mu \nu} \partial_{\mu} \phi  \partial_{\nu} \phi
 + 2 m_{\phi}^2 \phi^2 \right] \nonumber \\
 &+ S_m \left[\psi_m, A^2(\phi) g_{\mu \nu} \right]
\end{align}
where $g_{\mu\nu}$ is the metric, $\phi$ is the scalar field, $m_{\phi}$ is 
the coupling parameter for the mass potential and $dV=d^4x \sqrt{-g}$. 
$S_m$ is the matter action, $\psi_m$
denoting the matter degrees of freedom. Without the unusual non-minimal matter coupling,
this action simply represents a massive scalar in general relativity, not any \emph{alternative}
theory of gravitation. However, 
$\psi_m$ does not couple directly to the metric $g_{\mu\nu}$, but to a conformally scaled version
$\tilde{g}_{\mu\nu} = A^2(\phi) g_{\mu\nu}$. This non-minimal coupling breaks the equivalence principle, and
differentiates this theory from GR.
$g_{\mu\nu}$  and $\tilde{g}_{\mu\nu}$ define the so-called Jordan and Einstein frames respectively.
In the following, Jordan frame variables are denoted using tildes to distinguish them
from those in the Einstein frame which are bare. 

Let us consider $A(\phi)=e^{\beta \phi^2/2}$ where $\beta$ is a constant parameter of the theory.
$\phi=0$ is a trivial solution of the theory and is
equivalent to GR. The main point of spontaneous scalarization is that this trivial solution
is not necessarily stable, and there is a stable configuration with $\phi \neq 0$ under certain
circumstances. That is, any small deviation from $\phi=0$ grows and leads to a non-trivial solution.
The reason for the initial growth can be readily seen in the equation of motion (EOM)
\begin{align} \label{scalar_eom}
  \Box_g \phi &= \left( - 8 \pi A^4 \frac{d\left( \ln A(\phi) \right)}{d(\phi^2)} \tilde{T} + m^2_\phi \right)\phi \nonumber\\
  &\approx  ( - 4 \pi \beta \tilde{T} + m^2_\phi )\phi 
\end{align}
where $\tilde{T}$ is the trace of the matter stress-energy tensor in the Jordan frame, and we linearize the
equation in the second line. Hence, the effect of the non-minimal coupling is the introduction of
a mass-square-like term $- 4 \pi \beta \tilde{T}$. Let us consider the spacetime of a 
single star consisting of a perfect fluid which is not heavily relativistic. In such a case, 
$\tilde{T}=-\tilde{\rho}+3\tilde{p} \approx -\rho <0$. Thus, there are always negative values of 
$\beta$\footnote{Positive $\beta$ can also lead to spontaneous scalarization whose
behaviour can be different from $\beta<0$~\cite{Mendes:2016fby}. We will concentrate on the
$\beta<0$ case here.} 
that lead to
$ - 4 \pi \beta \tilde{T} + m^2_\phi <0$, and the coefficient of $\phi$ acts as the square of an imaginary mass term,
which causes exponential growth in time rather than oscillations. 
A field with imaginary mass and the consequent unstable behaviour is called a \emph{tachyon}. 
This exponential growth is the underlying reason for the instability of the $\phi=0$ solution, i.e.
even though $\phi=0$ is a solution in the presence of matter, it is not stable.

An instability is intriguing, but an indefinitely growing field cannot provide us with a well-behaved theory. We desire
this instability to saturate at some point and lead to a stable solution with finite $\phi \neq 0$, i.e. a star surrounded by a 
finite scalar field cloud. All fully non-linear calculations so far show that the instability is indeed regularized, leading to
stable scalar clouds~\cite{PhysRevLett.70.2220,Ramazanoglu:2016kul}. 
The main reason for the regularization of the tachyon is the nonlinear terms in the full EOM that 
we ignored in the linearized version, which make the potential term positive for large values
of $\phi$~\cite{Ramazanoglu:2016kul}. 
The linearized EOM is tachyonic only
at parts of the spacetime where the stellar matter exists, $\tilde{T} \neq 0$, and
it is a usual massive scalar outside per Eq.~\ref{scalar_eom}. 
Hence, the tachyonic instability is confined to a finite volume of the space, which possibly plays a role
in regularization as well. The size of the region the tachyon exists also dictates what 
values of $\beta$ allow spontaneous growth, which is the reason this phenomena is
only possible for compact stars~\cite{Ramazanoglu:2016kul}.

Solutions of spontaneously scalarized stars
and their fully non-linear dynamics has been studied by many
groups~\cite{2013PhRvD..87h1506B,2014PhRvD..89h4005S,Gerosa:2016fri,PhysRevD.93.124004}.
One attractive feature of spontaneous scalarization is that scalar field amplitudes can achieve very high
values so that the deviations from GR are not perturbative, but of order unity. Thus, these theories
can be easily detected with observations on strongly gravitating systems, such as binary mergers.
More detail about the allowed values of the $\beta-m_\phi$ parameters
and connections to observations can be found in~\cite{Ramazanoglu:2016kul}.

Lastly, we note that these results are not specific to a Gaussian, and 
can be extended to any function
with an ``inverse parabola'' expansion $A(\phi) = 1 -|\beta| \phi^2 + \ldots$.

\section{Ghost-based Spontaneous Scalarization}
Spontaneous scalarization can be seen as the introduction of a finite volume instability to
GR, which is eventually regulated by non-linear effects. We have presented
the specific case of a tachyon, but the arguments for the eventual regularization of the scalar
field can be also valid for other instabilities. In this section we will 
show that scalarization occurs for the instability of a ghost field as well.
While a tachyon occurs when there is an effective potential term with no lower
bound, a ghost occurs when there is an effective kinetic term with no lower bound. A typical example is a
classical Hamiltonian with a quadratic momentum term that has a negative coefficient. 
Energy can decrease indefinitely by increasing the magnitude of the momentum, leading to an instability.

The tachyon of the last section was introduced by 
a conformal scaling function $A$ which resembles the potential term in the scalar action,
i.e it was quadratic in $\phi$. This eventually leads to a mass term in the EOM 
which possesses the ``wrong'' sign, and competes with $\mphi$. 
Similarly, a ghost can be introduced by a scaling function that resembles the kinetic term:
\begin{align}\label{st_action_ghost}
 \frac{1}{16\pi} &\int dV R -  \frac{1}{16\pi} \int dV \left[ 2g^{\mu \nu} \partial_{\mu} \phi  \partial_{\nu} \phi
 + 2 m_{\phi}^2 \phi^2 \right] \nonumber \\
 &+ S_m \left[\psi_m, A_{\partial}^2(\eta) g_{\mu \nu} \right] \ , \ \eta \equiv g^{\mu\nu}\partial_\mu \phi \partial_\nu \phi
\end{align}
where $A_{\partial}$ is a conformal scaling function in analogy to Eq.~\ref{st_action}, but depends on the derivatives
of the scalar field rather than the field itself. We will continue to use the notation of the Jordan and 
Einstein frames.
Variations of the action lead to the EOMs
\begin{align}\label{ghost_EOM}
 R_{\mu\nu} - 8\pi \left( T_{\mu\nu} -\frac{1}{2} g_{\mu\nu} T \right) - 2 \partial_{\mu} \phi \partial_{\nu} \phi
 - \mphi^2 \phi^2 g_{\mu\nu} &=0 \nonumber \\
\nabla_\mu \left[(-8\pi\tilde{T}A_\partial^4 \alpha_\partial+1) \nabla^\mu \phi\right] - m_\phi^2 \phi & = 0 \ , 
\end{align}
where $\alpha_\partial = d(\ln A_\partial) /d\eta$. Reorganizing the scalar EOM
and using $A_\partial(\eta)=e^{\beta_\partial \eta/2}$ in analogy to $A$, the
linearized equation around $\phi=0$ becomes
\begin{equation}\label{ghost_EOM2}
 \left(-4\pi A_{\partial}^4 \beta_{\partial} \tilde{T} +1 \right)\Box_g \phi \approx \mphi^2 \phi 
 +4\pi A_{\partial}^4 \beta_{\partial} \nabla_\mu \tilde{T}\ \nabla^\mu\phi  \ .
\end{equation}
As long as $\tilde{T}$ is nonzero there are always values of $\beta_\partial$ that make the coefficient of
$\Box_g \phi$ negative. This is an exact analogue of spontaneous
scalarization that we studied in the previous section. The kinetic term
has the ``wrong'' sign now, which is the main feature of a ghost. $\phi$ tends to grow similarly to
a tachyon, hence is capable of growing arbitrarily small deviations from $\phi=0$.
This is still spontaneous scalarization, albeit arising from a different instability.
We will call this newly devised mechanism \emph{ghost-based spontaneous scalarization} to distinguish it from
the previous case which we will call \emph{tachyon-based spontaneous scalarization}. 
We can rearrange Eq.~\ref{ghost_EOM2} and define an effective mass
\begin{align}
m_{\rm eff}^2 \equiv  (-4\pi A_\partial^4 \beta_{\partial} \tilde{T} +1 )^{-1} \mphi^2
\end{align}
where $m_{\rm eff}^2<0$ when the field behaves like a ghost. Hence, a ghost and a tachyon behave
similarly at the level of the EOM which is another way to see the instability of the ghost. 

The next crucial requirement for a successful spontaneous scalarization theory is the eventual
regularization of the instability. This is accomplished by nonlinear terms,
and perhaps helped by the existence of the
unstable EOM only in a finite volume for the case of the tachyon. 
The same mechanisms are also present for the ghost-based version as the
form and especially the sign of $m_{\rm eff}$ is closely related to the case of the tachyon.
Thus, we also expect the ghost to be
eventually regularized, causing a stable nonzero scalar field configuration. To see
this explicitly, we calculate scalarized NS solutions of this theory.

We can derive the TOV-like equations for a static, spherically symmetric star 
from the metric ansatz
\begin{equation}
g_{\mu\nu} dx^{\mu} dx^{\nu} = -e^{\nu(r)} dt^2 + \frac{dr^2}{1-2\mu(r)/r} + r^2 d\Omega^2.
\end{equation}
We assume NS matter to be a perfect fluid with
$\tilde{T}^{\mu\nu}=(\rt+\pt)\tilde{u}^{\mu}\tilde{u}^{\nu}+\pt \tilde{g}^{\mu\nu}$
where the energy density $\rt$, pressure $\pt$, and components of the fluid 4-velocity $\tilde{u}^{\alpha}$
only depend on the radial coordinate $r$.
EOMs reduce to the following set of ordinary differential equations (ODEs):
\begin{align}
\label{tov}
 x' &= \left(8\pi A_\partial^4 \rt + \mphi^2 \phi^2 + \chi^2 \right) r- x/r \nonumber \\
 \nu' &= \left[8\pi A_\partial^4\pt -\mphi^2\phi^2 +(2\gamma-1)\chi^2  +x/r^2
 \right]r/(1-x)  \nonumber \\
 \phi'&= \chi/\sqrt{1-x} \nonumber \\
C_\chi \chi'&= -\left[C_\nu \nu'/2+2/r\right]  \chi+ \gamma^{-1} m_{\phi}^2 \phi/\sqrt{1-x} \nonumber \\
 \pt' &= -(\rt+\pt)\left[ \nu'/2+2\alpha_\partial \chi \chi' \right]
\end{align}
where $x=2\mu/r$, $\chi = \psi\sqrt{1-x}$, $\gamma = 1-8\pi A_\partial^4 \alpha_\partial  \tilde{T}$,
$'$ denotes a derivative with respect to $r$ and
\begin{align}
C_\chi &= 1-16 \pi A_\partial^4\gamma^{-1}  \chi^2 \nonumber \\
&\phantom{=1}\times [(4\alpha_\partial^2 +\alpha_\partial') (-\rt+3\pt) 
 -\alpha_\partial^2 (-\rt'+3)(\rt+\pt)] \nonumber \\
C_\nu &=1+8\pi A_\partial^4 \alpha_\partial \gamma^{-1}(-\rt'+3)(\rt+\pt) \nonumber
\end{align}
with $\rt' \equiv d\rt /d\pt$ and $\alpha_\partial' \equiv d \alpha_\partial/d\eta$.
This system of equations is completed by an equation of state (EOS) $\rt=\rt(\pt)$. 

The sign of $\gamma$ determines the ghost nature of $\phi$ in 
Eq.~\ref{ghost_EOM2}. $\gamma<0$ has to be satisfied in some region of the NS if a ghost is to
exist at all. $\gamma=1$ outside the NS where $\tilde{T}=0$, so any physically sensible solution that
has continuous TOV variables dictates $\gamma=0$ at some radius $r_\star$ inside the star.
This is potentially troublesome since $\chi'$ and $\tilde{p}'$ diverge at this point due to the 
$m_{\rm eff}^2 = \gamma^{-1} \mphi^2$ term. However, these divergences in
the derivatives do not imply that $\chi$ or $\tilde{p}$ themselves diverge, rather they have cusps.
More specifically, they behave as $C_1 +C_2|r-r_\star|^n$ near $r_\star$ with $0<n<1$
and $C_1,C_2$ constants.

A thorough understanding of the stability of this solution is only possible with time evolution which
is left to future studies.
We leave time evolution and a more detailed study of the effects of different NS EOS
and $m_\phi-\beta_\partial$ parameters to future studies.
The effect of the cusp on stationary NSs and during time evolution are especially interesting 
to investigate, in addition to order-of-unity deviations from GR that is the 
hallmark of spontaneous scalarization theories in general.\footnote{We presented
the numerical solution of a scalarized neutron star in the published version of this
study, however the equations of motion we used for the solution were
missing some terms which are corrected in the current version.
Hence, the published numerical solution is not accurate. Nevertheless it features
the most striking aspect of the star structure, the cusp.}

\section{Ghost-based Spontaneous Tensorization}
Just as the type of instability is not 
critical for spontaneous growth, recent studies have shown that the type of the field that carries the
instability is not critical either, e.g. one can replace the scalar with a vector field and
still observe spontaneous growth. This has been recently studied for
a tachyon-like instability living on a vector field~\cite{Ramazanoglu:2017xbl}
\begin{align}\label{action_vt}
 \frac{1}{16\pi} &\int d^4x \sqrt{-g} R -  \frac{1}{16\pi} \int d^4x \sqrt{-g} \left[ F^{\mu\nu} F_{\mu\nu}  +2m_X^2 X^\mu X_\mu \right] \nonumber \\
 &+ S_m \left[\psi_m, A_X^2(\eta) g_{\mu \nu} \right], \ \eta =g^{\mu\nu}X_\mu X_\nu
\end{align}
where $F_{\mu\nu} = \nabla_\mu X_\nu -\nabla_\nu X_\mu$,
and the matter terms are similar to Eq.~\ref{st_action}. 
Analogy to spontaneous scalarization is manifest in the EOM
\begin{align} \label{eom_vt}
\nabla_\rho F^{\rho \mu}& = \left(-8\pi A_X^4 \alpha_X \tilde{T} +m_X^2 \right) X^\mu 
\end{align}
where $\alpha_X(\eta) = d (\ln A_X(\eta))/d(\eta)$ and $\tilde{T}_{\mu \nu}$ is the stress-energy tensor 
in the Jordan frame. It can be shown that the vector field can spontaneously grow for appropriate
choices of $A_X$ leading to non-trivial vector field clouds around neutron stars.
This generalization of spontaneous growth to vectors is
called \emph{spontaneous vectorization}, and the
general term for spontaneous growth of any type of fields is named
\emph{spontaneous tensorization}. Close similarity of spontaneous vectorization in
Eq.~\ref{action_vt} to spontaneous scalarization strongly suggests
the stability of vectorized neutron stars, but no time evolution has been performed to 
explicitly demonstrate this~\cite{Ramazanoglu:2017xbl}.

The next natural step is combining the two ideas, that is
generalizing both the type of instability and the field that carries the instability. The simplest such 
generalization is \emph{ghost-based spontaneous
vectorization}. The form of the action is straightforward when we replace the scalar with a vector
and remember that $A_\partial$ resembled the form of the kinetic term in ghost-based
spontaneous scalarization
\begin{align}\label{action_vt_ghost}
 \frac{1}{16\pi} &\int dV R -  \frac{1}{16\pi} \int dV \left[ F^{\mu\nu} F_{\mu\nu}  +2m_X^2 X^\mu X_\mu \right] \nonumber \\
 &+ S_m \left[\psi_m, A_F^2(\eta_F) g_{\mu \nu} \right], \ \eta_F =F^{\mu\nu} F_{\mu\nu}/2 \ .
\end{align}
The Jordan frame
is associated with the metric $\tilde{g}_{\mu \nu}=A_F^2 g_{\mu \nu}$. EOM for the vector is
\begin{align} \label{eom_vt_ghost}
\nabla_\rho[(-8\pi A_F^4 \alpha_F \tilde{T}\ +1) F^{\rho \mu}]& = m_X^2 X^\mu 
\end{align}
where $\alpha_F \equiv d \left(\ln A_F(\eta_F)\right)/d\eta_F$. This is a vector ghost 
for appropriate functions $ \alpha_F$, in exact analogy to ghost-based spontaneous 
scalarization. All our consequent arguments about the eventual regularization and astrophysical
relevance of the scalar ghost can also be repeated for the vector without significant obstacles. We
will not attempt a detailed study of these issues here, but we once more see that spontaneous
growth and instability regularization is relevant well beyond tachyonic instabilities of scalar fields.
Establishing the stability of any solution in this theory again awaits time evolution studies.

After scalars and vectors, the next target for spontaneous
growth in the tensor hierarchy is a rank-$2$ tensor representing a spin-$2$ particle.
Let us summarize how we have been constructing theories of spontaneous growth.
First, we start with the Einstein-Hilbert action and a minimally
coupled field (scalar or vector), $S_{EH} +S_{field}$,
hence everything is under the umbrella of GR. Second, we finalize
the construction by adding matter fields that are nonminimally coupled to the Einstein-frame metric 
by simply scaling the metric in the matter coupling by a function of the field or its derivative.
\begin{align} \label{general_structre}
S = S_{EH} +S_{field}+ S_m \left[\psi_m, A^2(field) g_{\mu \nu} \right]
\end{align}

A theory of spontaneously growing rank-$2$ tensors is less
straightforward to build, and the conventional reasoning for this hardship is as follows.
The Einstein frame metric $g_{\mu\nu}$ itself is a spin-$2$ field,
and theories of interacting spin-$2$ particles including massive spin-$2$ fields have historically been
considered to be ill-behaved due to their inclusion of ghosts~\cite{deRham:2014zqa}.
Note that the ghosts that appear in these
theories, let us call them \emph{unwanted ghosts}, 
are not similar to the ghosts that we intentionally introduced to instigate spontaneous growth in the previous sections.
Unwanted ghosts appear as a side effect of the interactions, behave as ghosts in all parts of the spacetime,
and most importantly do not eventually regularize. Hence, even the first part of our construction,
$S_{EH} +S_{field}$, fails when the field is spin-$2$ (a second metric). 

Even though the history of interacting spin-$2$ fields does not look promising for spontaneous growth
at first, ghost-free theories
have been successfully introduced recently, including minimal matter 
coupling~\cite{deRham:2014zqa,PhysRevLett.106.231101,Hassan2012}. Schematically, their
action is in the form
\begin{align} \label{action_bigravity}
S = S_{EH}(g) + S_{EH}(f) + S_{int}(f, g) 
+ S_m \left[\psi_m, g_{\mu \nu} \right]
\end{align}
where we have two separate Einstein-Hilbert actions for two metrics $f$ and $g$,
an interaction piece which has a specific form to eliminate any ghosts while making the
fields massive, and finally a minimal matter coupling to one of the metrics.
This renews our hope to have spontaneous growth for spin-$2$ fields. 
By following the logic for the lower spin fields, in Eq.~\ref{action_bigravity}
we simply change the matter coupling into
a nonminimal one with a conformal scaling
\begin{align} \label{nonminimal_matter_coupling}
S_m \left[\psi_m, g_{\mu \nu} \right] \to S_m \left[\psi_m, A_{spin-2}^2(f,g)\ g_{\mu \nu} \right] \ .
\end{align}
An appropriate choice of $A_{spin-2}$, as in the lower spin cases, can induce 
an instability in the presence of NSs. If this instability is eventually regularized, 
we can have a well-behaved alternative gravity theory with
spontaneously growing spin-$2$ fields.

The above ideas give us a roadmap to investigate spontaneous scalarizarion of
spin-$2$ fields, but there is a major obstacle on this path: 
Non-minimal matter couplings such as Eq.~\ref{nonminimal_matter_coupling}
in bigravity theories are known to
reintroduce the unwanted ghosts in many cases~\cite{deRham:2014fha}. 
This means in addition to the instability we
intentionally introduce for spontaneous growth, the unwanted ghost might also reappear.
 Hence the first issue to check is whether the unwanted 
ghost is reintroduced. If it is not, then we have a novel well-behaved alternative theory of
gravitation. Such a theory is interesting for connecting the idea of massive gravity
and bigravity with spontaneous growth. If the ghost is reintroduced and the theory is ill-behaved,
this roadmap has to be abandoned.

We want to mention a third, more speculative possibility. Even if the unwanted 
ghost reappears at the linear level, could it also be regularized by nonlinear effects
similarly to the instability directly arising from the nonminimal matter coupling?
Irrespective of the details, our proposed spontaneously 
growing rank-$2$ field theory behaves identically to the well-behaved, ghost free theory in vacuum
where the matter coupling is irrelevant. Thus, when considering the spacetime of an isolated NS, 
the unwanted ghost is confined within the volume of the star, just like the instability
that is intentionally introduced to cause spontaneous growth (whether it is a tachyon or a ghost).
If nonlinear terms behave as they behave in spontaneous scalarization, this finite volume
unwanted ghost might also be regularized. This is a rather interesting possibility, and the 
nonlinear behavior of the unwanted ghost can be analyzed using techniques
in massive gravity literature~\cite{deRham:2014zqa,PhysRevLett.106.231101,Hassan2012}.
Nevertheless, we should emphasize that all our discussions on spontaneous
growth of spin-$2$ fields are speculation before any concrete studies to test them.

\section{Discussion}
The core idea of this study is the fact that instabilities in gravity theories can be regularized by nonlinear
effects, especially when higher order interactions are present and the instabilities are confined to a finite volume
in space. This can render them harmless and sometimes
even useful. This has been known in the case of spontaneous scalarization theories where a tachyonic instability
is regularized to form stable scalar clouds around NSs. 
The large deviations from GR that are caused by these clouds are an attractive target for 
gravitational wave observations and other tests of strong gravity.
Since there is nothing specific about tachyons in terms of their growth and regularization,
it is natural to expect a similar result for other instabilities. 
We verify this idea by showing that ghost instabilities also lead to nontrivial scalarized NS solutions.
Furthermore, the type of the field that possesses the instability, scalar vs vector,
is not crucial either. The regularization mechanism of
spontaneous scalarization seems to be powerful over many forms of instabilities, showing
that tachyon-based spontaneous scalarization is only one member of a vast family of 
spontaneous growth theories. However, 
the stability of the solutions for these newly identified theories
remains to be studied.

Following our line of thought brought us to a theory of spontaneous growth of spin-$2$
fields, which also makes a strong connection to massive gravity and bigravity. Interacting
spin-$2$ fields are plagued by unregularized ghosts that render them unphysical.
Even though these ghosts can be removed by recent breakthroughs in the field, nonminimal
matter couplings, such as the ones we introduce for spontaneous growth,
are known to bring them back in many situations. 
We discussed the three main possibilities for a proposed theory of spontaneously growing spin-$2$ fields.
First, the nonminimal matter coupling we introduce for spontaneous growth
somehow does not reintroduce the unwanted ghost,
and our idea is successful. Second, the ghost is reintroduced and is ill-behaved in the most
general case (in the nonlinear regime), meaning our proposal fails.
Third, even if the unwanted ghost is reintroduced at the linear level,
it is regularized at the nonlinear level by mechanisms similar to those we discussed for spontaneous growth, 
still leading to a well-behaved theory.
The last possibility might present an opportunity to test the power of the regularization mechanisms on
an instability that is not purposefully introduced.
We again emphasize that our discussion of spin-$2$ fields, especially the
third possibility, are not beyond the level of speculation before we concretely
study them in future work

Even though we concentrated on instabilities arising from conformal scalings of the matter coupling,
our ideas can be applied to other couplings as well. For the
specific case of scalar fields in gravity, possible terms in the Lagrangian have been classified by Horndeski and
recent extensions of his work~\cite{Horndeski1974,PhysRevLett.114.211101}. 
In addition to the conformal scalings in the non-minimal matter coupling,
more general options like disformal and extended
disformal couplings might also lead to spontaneous growth, and provide novel examples
of instability regularization ~\cite{Zumalacarregui:2013pma}.
Overall, understanding of instabilities provides valuable insight for both formal aspects of gravity 
theories and observable signatures of alternative theories.

\acknowledgments
Numerical calculations were performed on the Perseus
cluster at Princeton University and L\"{u}fer cluster at Ko\c{c} University.
Many ideas we present here are the result of our stimulating discussions with
Bayram Tekin, to whom we are deeply thankful. We are also grateful to Metin Murado\u{g}lu for
his help with numerical methods for stiff ODEs.

\bibliography{references_all}

\end{document}